\documentclass[trackchanges]{aastex701}
\usepackage{multirow}
\usepackage{booktabs}

\newcommand{\kms}{km\,s$^{-1}$}

\newcommand{\Msunyr}{\,M$_{\odot}$yr$^{-1}$}

\begin{document}

\title{Oxygen isotopes reveal low-mass star dominance in the Small Magellanic Cloud}

\author[orcid=0000-0002-3866-414X,sname='Gong']{Yan Gong}
%\altaffiliation{Kitt Peak National Observatory}
\affiliation{Purple Mountain Observatory, Chinese Academy of Sciences, 10 Yuanhua Road, Nanjing 210023, People's Republic of China}
\affiliation{Max-Planck-Institut f{\"u}r Radioastronomie, Auf dem H{\"u}gel 69, D-53121 Bonn, Germany}
\email[show]{ygong@pmo.ac.cn}  

\author[orcid=0000-0002-7299-2876,sname='Zhang']{Zhi-yu Zhang} 
\affiliation{School of Astronomy and Space Science, Nanjing University, Nanjing 210023, People's Republic of China}
\affiliation{Key Laboratory of Modern Astronomy and Astrophysics, Nanjing University, Ministry of Education, Nanjing 210023, People's Republic of China}
\email{zzhang@nju.edu.cn}

\author[orcid=0000-0002-7495-4005,sname='Henkel']{Christian Henkel} 
\affiliation{Max-Planck-Institut f{\"u}r Radioastronomie, Auf dem H{\"u}gel 69, D-53121 Bonn, Germany}
\email{chenkel@mpifr-bonn.mpg.de}

\author[orcid=0000-0002-3925-9365,sname='Chen']{C.-H. Rosie Chen} 
\affiliation{Max-Planck-Institut f{\"u}r Radioastronomie, Auf dem H{\"u}gel 69, D-53121 Bonn, Germany}
\email{rchen@mpifr-bonn.mpg.de}

\author[orcid=0000-0002-3599-6608,sname='Yang']{Wenjin Yang} 
\affiliation{School of Astronomy and Space Science, Nanjing University, Nanjing 210023, People's Republic of China}
\affiliation{Key Laboratory of Modern Astronomy and Astrophysics, Nanjing University, Ministry of Education, Nanjing 210023, People's Republic of China}
\email{wjyang@nju.edu.cn}
\affiliation{Max-Planck-Institut f{\"u}r Radioastronomie, Auf dem H{\"u}gel 69, D-53121 Bonn, Germany}

\author[orcid=0000-0002-4154-4309,sname='Chen']{Xindi Tang} 
\affiliation{Xinjiang Astronomical Observatory, Chinese Academy of Sciences, 830011 Urumqi, People's Republic of China}
\affiliation{Key Laboratory of Radio Astronomy, Chinese Academy of Sciences, 830011 Urumqi, People's Republic of China}
\email{tangxindi@xao.ac.cn}

\author[orcid=0000-0001-9162-2371,sname='Hunt']{Leslie K. Hunt} 
\affiliation{INAF – Osservatorio Astrofisico di Arcetri, Largo E. Fermi 5, 50125 Firenze, Italy}
\email{lesliekipphunt@gmail.com}

\author[orcid=0000-0003-4678-3939,sname='Weiss']{Axel Weiss} 
\affiliation{Max-Planck-Institut f{\"u}r Radioastronomie, Auf dem H{\"u}gel 69, D-53121 Bonn, Germany}
\email{aweiss@mpifr-bonn.mpg.de}

\author[orcid=0000-0003-0933-7112,sname='Chen']{Gang Wu} 
\affiliation{Xinjiang Astronomical Observatory, Chinese Academy of Sciences, 830011 Urumqi, People's Republic of China}
\affiliation{Key Laboratory of Radio Astronomy, Chinese Academy of Sciences, 830011 Urumqi, People's Republic of China}
\email{wug@xao.ac.cn}

\author[orcid=0000-0001-5574-0549,sname='Chen']{Yaoting Yan} 
\affiliation{Max-Planck-Institut f{\"u}r Radioastronomie, Auf dem H{\"u}gel 69, D-53121 Bonn, Germany}
\email{yyan@mpifr-bonn.mpg.de}

\author[orcid=0000-0002-3689-2392,sname='Grishunin']{Konstantin Grishunin} 
\affiliation{Max-Planck-Institut f{\"u}r Radioastronomie, Auf dem H{\"u}gel 69, D-53121 Bonn, Germany}
\email{kgrishunin@mpifr-bonn.mpg.de}

\author[orcid=0000-0001-6459-0669,sname='Menten']{Karl M. Menten} 
\altaffiliation{Deceased. With gratitude to Prof. Karl M. Menten, our beloved friend, mentor, and collaborator, whose full support and trust always propelled our research forward.}
\affiliation{Max-Planck-Institut f{\"u}r Radioastronomie, Auf dem H{\"u}gel 69, D-53121 Bonn, Germany}
\email{kmenten@mpifr-bonn.mpg.de}

%\collaboration{all}{The Terra Mater collaboration}

%% Use the \collaboration command to identify collaborations. This command
%% takes an optional argument that is either a number or the word "all"
%% which tells the compiler how many of the authors above the command to
%% show. For example "\collaboration[all]{(DELVE Collaboration)}" wil include
%% all the authors above this command.
%%
%% Mark off the abstract in the ``abstract'' environment. 
\begin{abstract}
Oxygen isotope abundances and their ratios are fingerprints of stellar evolution and therefore provide a powerful tool in tracing the enrichment history of galaxies. However, their behavior in low-metallicity dwarf galaxies remains largely unexplored. The Small Magellanic Cloud (SMC), a nearby analog of young high-redshift galaxies, offers an ideal laboratory to investigate this regime. Using the Atacama Compact Array, we observed the $J=2\to 1$ transitions of $^{12}$CO, $^{13}$CO, C$^{18}$O, and C$^{17}$O from the massive star-forming region LIRS~36 (aka N12A), achieving the first detection of C$^{17}$O in the SMC. This detection enables the first direct measurement of the $^{18}$O/$^{17}$O abundance ratio of 0.87$\pm$0.26 in this galaxy, substantially lower than all values in the literature, including molecular clouds in the Milky Way and other galaxies. Such a low ratio of $^{18}$O/$^{17}$O, together with a high $^{13}$CO/C$^{18}$O ratio, indicates chemical enrichment dominated by low-mass stars, consistent with the observed paucity of high-mass stars in the SMC. We suggest that the SMC is governed by a top-light integrated galaxy-wide initial mass function, predicted by the SMC’s persistently low star-formation activities. 
\end{abstract}

%% Keywords should appear after the \end{abstract} command. 
%% The AAS Journals now uses Unified Astronomy Thesaurus (UAT) concepts:
%% https://astrothesaurus.org
%% You will be asked to selected these concepts during the submission process
%% but this old "keyword" functionality is maintained in case authors want
%% to include these concepts in their preprints.
%%
%% You can use the \uat command to link your UAT concepts back its source.
\keywords{\uat{Galaxies}{573} --- \uat{Galaxy chemical evolution}{580} --- \uat{Interstellar medium}{847} --- \uat{Interstellar molecules}{849} --- \uat{Small Magellanic Cloud}{1468}}

%% From the front matter, we move on to the body of the paper.
%% Sections are demarcated by \section and \subsection, respectively.
%% Observe the use of the LaTeX \label
%% command after the \subsection to give a symbolic KEY to the
%% subsection for cross-referencing in a \ref command.
%% You can use LaTeX's \ref and \label commands to keep track of
%% cross-references to sections, equations, tables, and figures.
%% That way, if you change the order of any elements, LaTeX will
%% automatically renumber them.

\section{Introduction}\label{sec.intro}
Isotopic abundance ratios serve as powerful diagnostics of stellar nucleosynthesis, providing insight into the impact of stellar yields, the composition of stellar populations, and the broader context of galactic chemical evolution \citep[e.g.,][]{1994ARA&A..32..191W,2017MNRAS.470..401R,2018Natur.558..260Z,2022A&ARv..30....7R,2023AA...670A..98Y,2023A&A...679L...6G,2024ApJ...970..136G}. Oxygen, the most abundant metal element in the Universe, exhibits isotopic diversity with rare isotopes. For example, $^{18}$O is primarily synthesized through $\rm ^{14}N(\alpha, \gamma)^{18}F(\beta^+)^{18}O$ reaction during helium
burning in massive stars, while $^{17}$O is produced in the hydrogen-burning zones of lower-mass stars via the cold or hot CNO cycles \citep[e.g.,][]{1994ARA&A..32..191W,2022A&ARv..30....7R}. Novae may represent an important source of $^{17}$O enrichment, as shown in Galactic chemical evolution simulations \citep{2003MNRAS.342..185R} and nova model calculations \citep{1998ApJ...494..680J}. Consequently, $^{17}$O is commonly injected into the interstellar medium (ISM) more slowly than $^{18}$O.

On the other hand, oxygen isotopes are not subject to chemical fractionation \citep[e.g.,][]{1984ApJ...277..581L}, since the scarcity of O$^{+}$ in molecular clouds make charge–exchange reactions negligible. Furthermore, the $^{18}$O/$^{17}$O isotope ratios can be directly determined by the optically thin transitions of C$^{18}$O and C$^{17}$O \citep[e.g.,][]{2008A&A...487..237W,2020ApJS..249....6Z,2023MNRAS.522..559O}. Isotope-selective photodissociation is unlikely to significantly affect these ratios, as both CO isotopologues exhibit similar self-shielding properties \citep[e.g.,][]{2009A&A...503..323V}. These properties make $^{18}$O/$^{17}$O isotope ratios a good probe of nuclear processing and metal enrichment in galaxies. 

Recent studies have revealed a tentative gradient in the Milky Way, with the $^{18}$O/$^{17}$O isotope ratio increasing from the Galactic center ($\sim$3) toward the outer Galaxy \citep[$\sim$5; e.g.,][]{2008A&A...487..237W,2020ApJS..249....6Z,2023MNRAS.522..559O,2023AA...670A..98Y}. This finding supports an inside-out formation scenario for the Galactic disk \citep[e.g.,][]{2001ApJ...554.1044C,2012A&A...540A..56P}. Beyond the Milky Way, $^{18}$O/$^{17}$O ratios have been measured to be $\sim$1.7 in the Large Magellanic Cloud (LMC) \citep{1998A&A...332..493H,2009ApJ...690..580W}, $\gtrsim$8 in nearby starburst galaxies \citep[e.g.,][]{1993AA...274..730H,2021AA...656A..46M}, and $>$10 in two intermediate redshift galaxies at $z =0.68$ and 0.89 \citep{1995A&A...303L..61C,2006A&A...458..417M,2023AA...674A.101M}. Despite these advances, determining the oxygen isotope ratios in very metal-poor galaxies ($Z\lesssim$0.2~$Z_{\odot}$), including local dwarf galaxies and low-metallicity galaxies at high redshift, is challenging, but essential to understand their role in the oxygen enrichment of the Universe \citep[e.g.,][]{2022Galax..10...11H}.

The Small Magellanic Cloud (SMC), at a distance of 62.1$\pm$1.9~kpc \citep{2014ApJ...780...59G}, is a nearby dwarf irregular galaxy with low star formation rates \citep{2004AJ....127.1531H}. Its gas-phase metallicity is characterized by an oxygen abundance of O/H$\sim 10^{-4}$ ($\sim$0.2~$Z_{\odot}$; e.g., \citealt{1997macl.book.....W}), while its nitrogen abundance is even more deficient at N/H$\sim 3.2\times10^{-6}$ \citep[e.g.,][]{1998RMxAC...7..202K,2022MNRAS.517.4497D}. These make the SMC the best testbed to study metal-poor environments in the Local Group and a unique prototype for understanding the properties of metal-poor galaxies which were ubiquitous in the early Universe. Precise measurements of the gas-phase $^{18}$O/$^{17}$O isotope ratios in the SMC will yield important insight into its chemical evolution and establish it as a benchmark for studies of extremely metal-poor galaxies.

Several attempts have been undertaken to measure the isotope ratios of elements in the SMC. However, the results are still inconclusive. For instance, \citet{1998A&A...330..901C} and \citet{1998A&A...332..493H} attempted to detect C$^{18}$O in the SMC, but were unsuccessful. Their lower limits of the $^{13}$CO/C$^{18}$O line intensity ratios are much higher than those observed in Galactic molecular clouds. This finding was later confirmed by ALMA observations \citep{2017ApJ...844...98M,2018Natur.558..260Z}, which revealed remarkably high intensity ratios of around 100. Similarly, \citet{1999A&A...344..817H} reported a non-detection of C$^{18}$O and C$^{17}$O. As a result, the oxygen isotope ratios in the SMC remain largely unconstrained. Therefore, we performed new ALMA observations toward a molecular cloud in SMC, LIRS~36 (aka N12A), to measure the C$^{18}$O and C$^{17}$O emission and their abundance ratios. The observations are described in Sect.~\ref{Sec:obs}, and the results are presented in Sect.~\ref{Sec:res}. We discuss the implications of our findings in Sect.~\ref{Sec:dis}, and provide a summary in Sect.~\ref{Sec:sum}.

%These findings highlight the need for further investigation of isotope ratios in the SMC to enhance our understanding of the oxygen enrichment of this nearby metal-deficient galaxy.

\section{Observations and data reduction}\label{Sec:obs}
Based on the recently completed APEX $^{12}$CO $J=3\to 2$ and $^{13}$CO $J=3\to 2$ survey (C.-H.~Rosie~Chen, in prep.), which covers 5.1 square degrees of molecular clouds in the SMC, LIRS~36 in the southwest bar is identified as the brightest position in $^{13}$CO $J=3\to 2$ emission (see also \citealt{2024A&A...687A..26S} for a smaller $^{12}$CO $J=3\to 2$ survey of the SMC Bar at a shallower depth). The observations have an angular resolution of 18\arcsec ($\sim$5.4~pc). This indicates that LIRS~36 has the highest $^{13}$CO (and therefore H$_{2}$) column density on the scale probed by the survey. Previous single-dish observations also suggest that LIRS~36 is one of the most prolific molecular line sources in the SMC \citep{1998A&A...330..901C,2016ApJ...822..105A}. Therefore, we selected LIRS~36 as the target to search for rare CO isotopologues. 

The observations of LIRS~36 (project code: 2023.1.01576.S, PI: Yan Gong) were conducted between 2024 June 24 and August 23 using the Atacama Compact Array \citep[ACA;][]{2009PASJ...61....1I}. The phase center of the observations was located at ($\alpha_{\rm J2000}$, $\delta_{\rm J2000}$)=(00$^{\rm h}$46$^{\rm m}$40\rlap{.}$^{\rm s}$8, $-$73$^{\circ}$06$^{\prime}$14\rlap{.}$^{\prime\prime}$4). Observations were performed using the ALMA Band 6 receivers, covering both spectral line and continuum emission with two frequency setups comprising 16 spectral windows in total. In this study, we focus on four spectral windows centered on the $^{12}$CO $J=2\to1$, $^{13}$CO $J=2\to1$, C$^{18}$O $J=2\to1$, and C$^{17}$O $J=2\to1$ transitions. The former three transitions were observed simultaneously, while the last line was observed independently. The spectral channel width was 244~kHz, corresponding to a velocity spacing of $\sim$0.3~\kms\,for the four transitions. The bandpass and flux calibrator was J2258$-$2758, while the phase calibrator was J0102$-$7546. The observations employed 7--10 antennas with baseline lengths of 8.9--48.9 m and a total of $\sim$60 hours of telescope time.

The raw data were calibrated using the ALMA pipeline of the Common Astronomy Software Applications (CASA) package \citep{2022PASP..134k4501C}. Subsequent imaging of the calibrated data was performed manually with CASA version 6.5.4. The image reconstruction was carried out using the ``tclean'' task, employing the H{\"o}gbom algorithm \citep{1974A&AS...15..417H} and Briggs weighting with a robust parameter of 0.5. The automated masking algorithm ``AUTO-MULTITHRESH'' was adopted in the task, with parameters set identically to those listed in Table~2 of \citet{2020PASP..132b4505K}. All cleaned images have a pixel size of 1\rlap{.}\arcsec1, corresponding linearly to approximately 1/5 of their synthesized beams. A primary beam correction was applied to all cleaned data. The observational details are summarized in Table~\ref{Tab:lin}. Based on the measurements of the calibrators, the flux calibration accuracy is estimated to be better than 10\%.

Total-power (TP) $^{12}$CO $J=2\to 1$, $^{13}$CO $J=2\to 1$, and C$^{18}$O $J=2\to 1$ observations were conducted in 2016 June and August (project code: 2015.1.00196.S; PI: Julia Roman-Duval). The target region was mapped using on-the-fly (OTF) mapping with Nyquist sampling, covering a 2$\arcmin \times$2$\arcmin$ rectangular region centered at ($\alpha_{\rm J2000}$, $\delta_{\rm J2000}$) = (00$^{\rm h}$46$^{\rm m}$41\rlap{.}$^{\rm s}$279, $-$73$^{\circ}$05$^{\prime}$77\rlap{.}$^{\prime\prime}$797). The single-dish data were reduced using the standard CASA pipeline. The half-power beam widths (HPBWs) are 29\rlap{.}$\arcsec$1, 30\rlap{.}$\arcsec$3, and 30\rlap{.}$\arcsec$4 for transitions of $^{12}$CO, $^{13}$CO, and C$^{18}$O, respectively. The typical 1-$\sigma$ noise levels per 244~kHz (i.e., 0.3~\kms) channel are $\sim$0.5~Jy~beam$^{-1}$ for both $^{12}$CO $J=2\to 1$ and $^{13}$CO $J=2\to 1$, and $\sim$~0.4~Jy~beam$^{-1}$ for C$^{18}$O $J=2\to 1$. In this study, we only adopt the $^{12}$CO $J=2\to 1$ and $^{13}$CO $J=2\to 1$ data to recover the missing flux in our ACA observations, as C$^{18}$O $J=2\to 1$ was not detected in the TP observations. The combination of the single-dish and ACA datasets was performed using the ``feather'' task in CASA. To reduce the different beam dilution effects, all data were convolved to a common circular beam of 7\rlap{.}\arcsec5 for subsequent analysis.

\begin{deluxetable*}{cccccccc}[htbp!]
\tabletypesize{\footnotesize}
\tablewidth{0pt}
\renewcommand\arraystretch{0.97}
\tablecaption{Summary of ACA Observations. \label{Tab:lin}}
\tablehead{
\colhead{Transition}  & \colhead{Frequency}  & \colhead{Date} & \colhead{PWV} & \colhead{$\theta_{\rm maj}\times \theta_{\rm min}$, $PA$} & \colhead{MRS}  & \colhead{$\sigma$} & \colhead{$\delta v$} \\
\colhead{}  & \colhead{(MHz)} & \colhead{} & \colhead{(mm)} & \colhead{(\arcsec$\times$\arcsec, \degr)} & \colhead{(\arcsec)} & \colhead{(mJy~beam$^{-1}$)} & \colhead{(\kms)}   \\
\colhead{(1)}  & \colhead{(2)} & \colhead{(3)} & \colhead{(4)} & \colhead{(5)} & \colhead{(6)} & \colhead{(7)} & \colhead{(8)} }
\startdata
$^{12}$CO $J=2\to 1$ & 230538.0000(5) & [2024-Jul-17, 2024-Aug-23] & 0.45--2.16 & 6.5$\times$5.7, $-$86 & 29.0 & 11 & 0.32 \\
$^{13}$CO $J=2\to 1$ & 220398.6842(1) & [2024-Jul-17, 2024-Aug-23] & 0.45--2.16 & 6.8$\times$6.8, 23    & 30.4 & 10 & 0.33 \\
C$^{18}$O $J=2\to 1$ & 219560.3577(3) & [2024-Jul-17, 2024-Aug-23] & 0.45--2.16 & 7.0$\times$6.1, $-$79 & 30.5 & 8 & 0.33 \\
C$^{17}$O $J=2\to 1$ & 224714.3858(3) & [2024-Jun-24, 2024-Jul-26] & 0.49--2.12 & 6.8$\times$5.9, $-$73 & 29.7 & 4 & 0.33 \\
\enddata
\tablecomments{(1) Transition. (2) Rest frequency. Uncertainties in the last decimal digit, suggested by the Cologne Database for Molecular Spectroscopy \citep[CDMS,][]{2016JMoSp.327...95E}, are given in parentheses. (3) Observing date. (4) Precipitable water vapor during the observations. (5) Restoring beam. (6) Maximum recoverable scale. (7) Noise level at the corresponding channel width. (8) Channel width in units of \kms.}
\end{deluxetable*}

\begin{deluxetable*}{ccccccccc}[!hbt]
\caption{Observed and physical properties of the transitions of CO and its isotopologues toward the peak position at the convolved beam of 7\rlap{.}\arcsec5.}\label{Tab:obs}
\tabletypesize{\small}
\centering
\tablehead{
\colhead{}  & \colhead{}  & \colhead{}  & \colhead{} & \colhead{} & \multicolumn{2}{c}{f=1}  & \multicolumn{2}{c}{f=0.34}     \\ [-2pt]
\cmidrule(lr){6-7} \cmidrule(lr){8-9} \\ [-14pt]
\colhead{Transition}             & \colhead{$v_{\rm lsr}$} & \colhead{$T_{\rm p}$}  & \colhead{$\Delta v$} & \colhead{$\int T_{\rm mb} {\rm d}v$} & \colhead{$T_{\rm ex}$}  & \colhead{$N_{\rm mol}$}  & \colhead{$T_{\rm ex}$}  & \colhead{$N_{\rm mol}$} \\ [-3pt]
\colhead{}                 & \colhead{(\kms)} & \colhead{(K)}     & \colhead{(\kms)}      & \colhead{(K~\kms)}          &  \colhead{(K)}  &  \colhead{(cm$^{-2}$)} & \colhead{(K)}  &  \colhead{(cm$^{-2}$)} \\ [-3pt]
\colhead{(1)}  & \colhead{(2)}   & \colhead{(3)}      & \colhead{(4)}         & \colhead{(5)}                &  \colhead{(6)}   & \colhead{(7)}          &  \colhead{(8)}  & \colhead{(9)}  }
\startdata
$^{12}$CO $J=2\to 1$ & 126.2$\pm$0.1 & 13.4$\pm$0.5 & 3.0$\pm$0.1 & 44.0$\pm$0.5 & 18.4$\pm$3 & \nodata & 46.5$\pm$8.2 & \nodata \\
$^{13}$CO $J=2\to 1$ & 126.1$\pm$0.1 & 3.88$\pm$0.04 & 2.1$\pm$0.1 & 8.88$\pm$0.02 &  \nodata   & (5.9$\pm$1.5)$\times 10^{15}$  & \nodata   & (2.5$\pm$0.6)$\times 10^{16}$  \\
C$^{18}$O $J=2\to 1$ & 125.9$\pm$0.1 & 0.055$\pm$0.004 & 1.6$\pm$0.1 & 0.099$\pm$0.005                       &  \nodata   & (5.4$\pm$1.1)$\times 10^{13}$  & \nodata   & (2.2$\pm$0.5)$\times 10^{14}$  \\
C$^{17}$O $J=2\to 1$ & 126.1$\pm$0.1 & 0.051$\pm$0.003 & 2.1$\pm$0.1 & 0.114$\pm$0.002                      &  \nodata   & (6.2$\pm$1.5)$\times 10^{13}$  & \nodata   & (2.5$\pm$0.6)$\times 10^{14}$  \\
\enddata
\tablecomments{(1) Transition. (2) Velocity centroid. (3) Peak brightness temperature. (4) Full
width at half maximum (FWHM) line width. (5) Integrated intensity. The values are obtained by integrating the velocity range of 105--145~\kms\,for $^{12}$CO ($J=2\to 1$) and 122--129~\kms\,for the other three transitions. (6)–(7) Excitation temperature and molecular column density assuming a beam filling factor of $f=1$. (8)–(9) Same as (6)–(7), but for $f=0.34$. The observed properties of $^{12}$CO ($2\to 1$) and $^{13}$CO ($2\to 1$) are derived from the ACA+TP combined datasets, whereas those of C$^{18}$O ($2 \to 1$) and C$^{17}$O ($2 \to 1$) are derived from the ACA-only datasets.}
\end{deluxetable*}

\section{Results}\label{Sec:res}

\begin{figure*}[!htbp]
\centering
\includegraphics[width = 0.9 \textwidth]{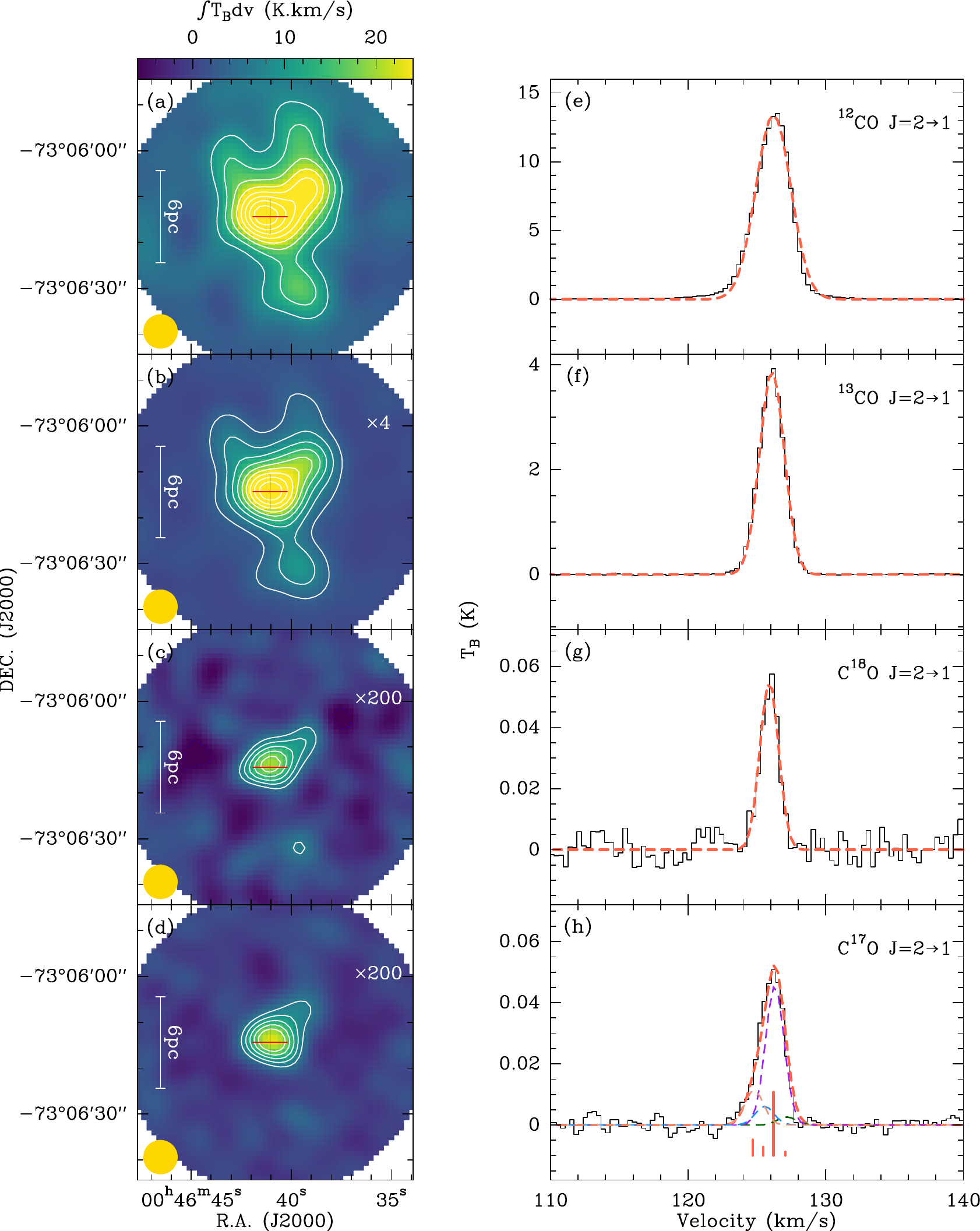}
\caption{{Spatial distributions and spectra of the four $J=2\to 1$ transitions of CO isotopologues. Panels (a)–(d) show the integrated-intensity maps of $^{12}$CO $J=2\to1$, $^{13}$CO $J=2\to1$, C$^{18}$O $J=2\to1$, and C$^{17}$O $J=2\to1$, respectively, overlaid with their corresponding contours. The velocity integration ranges are 105--145~\kms\,for $^{12}$CO $J=2\to1$ and 122--129~\kms\,for the other three transitions. Contours start at 9.0~K~\kms\,and increase in steps of 4.5~K~\kms\,in panel (a), at 0.89~K~\kms\,with steps of 0.89~K~\kms\,in panel (b), and at 0.03~K~\kms\,with steps of 0.015~K~\kms\,in panels (c)–(d). Panels (e)--(h) show the $^{12}$CO $J=2\to 1$, $^{13}$CO $J=2\to 1$, C$^{18}$O $J=2\to 1$, and C$^{17}$O $J=2\to 1$ spectra of the position indicated by the red plus in panels~(a)--(d). The red dashed curves in panels (e)--(h) show the Gaussian and hyperfine-structure fitting results, while the red vertical lines in panel (h) indicate the positions of the hyperfine-structure components. The $^{12}$CO $J=2\to1$ and $^{13}$CO $J=2\to1$ data are from ACA+TP combined observations, whereas the C$^{18}$O $J=2\to1$ and C$^{17}$O $J=2\to1$ data are from ACA-only observations. Flux-to-brightness temperature conversion factors are 2.4462, 2.2358, 2.2188, 2.3242~Jy~K$^{-1}$ at an angular resolution of 7.5$^{\prime\prime}$ for $^{12}$CO $J=2\to1$, $^{13}$CO $J=2\to1$, C$^{18}$O $J=2\to1$, and C$^{17}$O $J=2\to1$, respectively.}\label{Fig:N12A}}
\end{figure*}

Figures~\ref{Fig:N12A}a--\ref{Fig:N12A}d present the spatial distributions of the $J=2\to 1$ transitions of $^{12}$CO and its three isotopologues. All lines are detected and exhibit broadly similar morphologies, with emission peaking at nearly the center of the field of view and extending toward the northwest. The similar morphologies support the assumption that all transitions trace the same bulk molecular gas at the peak. Thanks to the high sensitivity and high angular resolution of our ACA observations, the C$^{17}$O $J=2\to1$ transition is detected with a signal-to-noise ratio of $>$10 (Fig.~\ref{Fig:N12A}d), marking the first solid detection of C$^{17}$O in the SMC.

The $^{13}$CO $J=2\to 1$ integrated intensity map peaks at ($\alpha_{\rm J2000}$, $\delta_{\rm J2000}$)=(00$^{\rm h}$46$^{\rm m}$41\rlap{.}$^{\rm s}$1, $-$73$^{\circ}$06$^{\prime}$14\rlap{.}$^{\prime\prime}$41). The observed spectra toward this position are shown in Figs.~\ref{Fig:N12A}e--\ref{Fig:N12A}h, and all exhibit single-peaked, Gaussian-like profiles. To mitigate the effects of potential missing flux (see Appendix~\ref{app.a} for further discussion), we extracted spectra at the emission peak and used them to derive the isotopic ratios. We performed a Gaussian fitting to each spectrum (indicated by red lines in Figs.~\ref{Fig:N12A}e--\ref{Fig:N12A}h), and the derived parameters are summarized in Table~\ref{Tab:obs}. The velocity centroids of the four transitions are consistent within the uncertainties and are in good agreement with previous studies \citep[e.g.,][]{1997A&A...317..548C,1998A&A...330..901C}. The $^{12}$CO $J=2\to 1$ line exhibits a broader FWHM than the other transitions, most likely due to opacity broadening, and also shows a weak blue-shifted wing not detected in the isotopologues. The integrated intensity ratio of $^{13}$CO/C$^{18}$O is 90$\pm$5, consistent with the high intensity ratios reported for N83C in the SMC \citep{2017ApJ...844...98M}. Such high ratios appear to be a common feature of the bright star-forming regions observed so far in the SMC. A galaxy-wide census of $^{13}$CO and C$^{18}$O emission would be required to assess whether this trend is representative of the global ISM in the SMC.

The C$^{17}$O $J=2\to 1$ line appears slightly broader than that of C$^{18}$O $J=2\to 1$ and shows a blue-shifted asymmetry profile, due to the hyperfine-structure splitting in C$^{17}$O $J=2\to 1$. In Fig.~\ref{Fig:N12A}g, we marked the corresponding frequencies and relative intensity of the hyperfine transitions. It is remarkable that the integrated intensity of C$^{17}$O $J=2\to 1$ exceeds that of C$^{18}$O $J=2\to 1$. To our knowledge, this is the first instance in which C$^{17}$O emission exhibits a higher integrated intensity
than C$^{18}$O emission in molecular clouds.

Assuming local thermodynamic equilibrium (LTE), optically thick $^{12}$CO $J=2\to1$ emission, and neglecting beam dilution (see Appendix~\ref{app.lte}), we derive an excitation temperature of $18.6 \pm 3.0$ K from the peak intensity
of the $^{12}$CO $J=2\to1$ line profile. Assuming that the four $J=2\to 1$ transitions have the same excitation temperature, we derive a peak optical depth of 0.34$\pm$0.02 for $^{13}$CO $J=2\to 1$. The optical depths of C$^{18}$O $J=2\to 1$ and C$^{17}$O $J=2\to 1$ are well below 0.01. Even if we assume a beam dilution factor of 0.34 (i.e., a source size of 5\rlap{.}\arcsec4; see Appendix~\ref{app.lte}), which increases the excitation temperature to 46.3$\pm$8.2~K, the optical depths of $^{13}$CO $J=2\to 1$, C$^{18}$O $J=2\to 1$, and C$^{17}$O $J=2\to 1$ stay almost unchanged. Hence, these three transitions can be considered optically thin for
our study.

Using the LTE approach within the Markov chain Monte Carlo (MCMC) framework (see Appendix~\ref{app.lte}) and adopting a 10\% uncertainty in flux density, we derive the $^{13}$CO, C$^{18}$O, and C$^{17}$O column densities, as listed in
Table~\ref{Tab:obs}. The corresponding isotope ratios were directly determined from these column densities. Neglecting the beam dilution effects, we obtain $^{13}$C/$^{18}$O=109.3$\pm$35.6, $^{13}$C/$^{17}$O=95.2$\pm$31.4, and $^{18}$O/$^{17}$O=0.87$\pm$0.26. The dependence of the derived isotope ratios on the assumed source sizes is negligible (see Appendix~\ref{app.lte}), indicating that these results are robust. Because the $^{12}$CO $J=2\to1$ transition is optically thick, the integrated intensity ratios of the $^{12}$CO
$J=2\to1$ line to the C$^{18}$O $J=2\to1$ and C$^{17}$O $J=2\to1$ lines yield only lower limits of $^{16}$O/$^{18}$O$\gtrsim$444 and
$^{16}$O/$^{17}$O$\gtrsim$386, respectively.

\section{Discussion}\label{Sec:dis}
%In this study, our measurements toward N12A provide the first determination of oxygen isotope ratios in the SMC. 
Table~\ref{Tab:ratio} compares oxygen isotope ratios with those measured in various environments. The $^{18}$O/$^{17}$O isotope ratio in the SMC is markedly lower than all previously reported values for molecular clouds in both extragalactic sources and the Milky Way. Given that $^{18}$O/$^{17}$O ratios are a valuable diagnostic for understanding galactic chemical enrichment, we explore the implications of this exceptionally low value below.

\begin{deluxetable}{ccccc}[!hbt]
\caption{Oxygen isotope ratios.}\label{Tab:ratio}
\tabletypesize{\small}
\centering
\tablehead{
\colhead{Region}            & \colhead{$^{18}$O/$^{17}$O} & \colhead{$^{16}$O/$^{18}$O} & \colhead{$^{16}$O/$^{17}$O} & \colhead{reference} }
\startdata
\multicolumn{5}{c}{Extragalactic sources} \\
\hline
SMC            & 0.87$\pm$0.26 & $\gtrsim$444 & $\gtrsim$386  & a \\
LMC            &  1.7$\pm$0.2  & 2000$\pm$250 & 3000$\pm$580  & b \\
NGC~253        &  8.7$\pm$1.2  & 48$\pm$5     &  400$\pm$40   & c \\
NGC~4945       &  6.4$\pm$0.3  & 200$\pm$50   & 1250$\pm$290  & d \\
M82            &  8$\pm$1      & $>$350       & $>$2800       & e, f\\
PKS 1830$-$211 &  11.5$\pm$0.5 & 65.3$\pm$0.7 &  751$\pm$34   & g \\    
NGC~3256       &  5--12        & \nodata      & \nodata       & h \\
Arp~220        &  $>$20        & \nodata      & \nodata       & i \\
M83            &  5.8--7.1     & \nodata      & \nodata       & j \\
\hline
\multicolumn{5}{c}{The Milky Way} \\
\hline
CMZ            &  3.4$\pm$0.1  &  260$\pm$50  & 890$\pm$160   & k \\
Inner disk     &  3.6$\pm$0.2  &  330$\pm$30  & 1180$\pm$130  & k \\
Local ISM      &  3.9$\pm$0.4  &  560$\pm$25  & 2180$\pm$240  & k \\
Outer Galaxy   &  4.8$\pm$0.6  &  630$\pm$140 & 3000$\pm$790  & k \\
Solar          &  5.5          & 490          & 2625          & l \\
\enddata
\tablecomments{$^a$This work; 
$^b$\citet{2009ApJ...690..580W}; $^c$\citet{2021AA...656A..46M}; $^d$\citet{2004AA...422..883W}; $^e$\citet{1993AA...274..730H}; 
$^f$\citet{2010AA...522A..62M};
$^g$\citet{2023AA...674A.101M};
$^h$\citet{2018ApJ...855...49H};
$^i$\citet{2015AA...579A.101A};
$^j$\citet{2018ApJ...855...49H}
$^k$\citet{2023AA...670A..98Y};
$^l$\citet{1989GeCoA..53..197A}.}
\end{deluxetable}

In the Milky Way, low $^{18}$O/$^{17}$O ratios (sometimes below unity) are commonly observed in the circumstellar envelopes of carbon-rich stars \citep[e.g.,][]{1994ARA&A..32..191W,2017A&A...600A..71D,2024Galax..12...70A}. The detection of such a low $^{18}$O/$^{17}$O ratio in the SMC therefore could indicate a substantial contribution from low- and intermediate-mass stars. This is consistent with the finding that, owing to its low metallicity, the SMC hosts a higher fraction of carbon-rich stars than both the Milky Way and the LMC \citep[e.g.,][]{1978Natur.271..638B,1980ApJ...242..938B,2005A&A...434..691M,2006ApJ...645.1118S}. 
At lower metallicities, reduced hot-bottom burning and more efficient third dredge-up episodes can help elevate C/O ratios
\citep[e.g.,][]{2013MNRAS.434..488M,2014PASA...31...30K,2018MNRAS.476..496F}, enhancing the prevalence and chemical impact of such stars.

The nucleosynthetic origins of the two oxygen isotopes are distinct: $^{17}$O is mainly synthesized in LIMs via $\rm ^{16}O(p, \gamma)^{17}F(\beta^+ \nu)^{17}O$, while $^{18}$O is produced predominantly in high-mass stars through $\rm
^{14}N(\alpha, \gamma)^{18}F(\beta^+)^{18}O$ \citep[e.g.,][]{1994ARA&A..32..191W,2003MNRAS.342..185R,2022A&ARv..30....7R}.
Consequently, an unusually low $^{18}$O/$^{17}$O ratio indicates chemical enrichment dominated by low- and intermediate-mass stellar populations, with little contribution from massive stars. This interpretation is consistent with the latest Gaia census of the SMC that finds the number of high-mass stars to be lower by a factor of 3--7 than expected from the low-mass population
\citep{2021A&A...646A.106S}, implying a steep slope of the high-mass end of the initial mass function (IMF). Such a top-light IMF may also reduce the number of intermediate-mass stars that dominate $^{14}$N production via the CNO cycle, yielding the markedly low nitrogen abundance observed in the SMC \citep[e.g.,][]{1998RMxAC...7..202K,2022MNRAS.517.4497D} and, in turn, suppressing $^{18}$O synthesis from $^{14}$N in stars. Together, the paucity of intermediate- and high-mass stars provides a coherent explanation for both the
$^{18}$O deficit and the low $^{18}$O/$^{17}$O ratio.

The abundances of elements and isotopes trace the cumulative chemical enrichment history of the SMC, extending back to the formation of its first-generation stars at least $\sim$10~Gyr ago \citep[e.g.,][]{2004AJ....127.1531H}. As oxygen isotopes originate from different stellar mass ranges, their relative abundance ratios are sensitive to the integrated galaxy-wide initial mass function
(IGIMF), which depends on star formation rates and metallicity \citep[e.g,][]{2017A&A...607A.126Y,2018A&A...620A..39J,2020A&A...637A..68Y,2023Natur.613..460L}. According to the reconstructed star formation history of the SMC \citep{2009AJ....138.1243H}, the overall star formation rates were low, of order
$<$0.3~\Msunyr\,over the past $\gtrsim$10~Gyr \citep[e.g.,][]{2004AJ....127.1531H}. Theoretical models predict that such low
star formation rates yield a top-light IGIMF, in which the formation of intermediate- and high-mass stars is strongly suppressed
\citep{2017A&A...607A.126Y,2018A&A...620A..39J}. The low $^{18}$O/$^{17}$O ratio that we measure is in line with this prediction, indicating long-term enrichment dominated by low-mass stars.

The ISM of the LMC appears to be well mixed, exhibiting nearly uniform isotope ratios across the galaxy \citep[e.g.,][]{1998A&A...332..493H}. In the SMC, large-scale mixing of nucleosynthetic products may also be expected, although this remains less well constrained. This expectation is supported by the similar $^{13}$CO/C$^{18}$O integrated intensity ratios observed in two different SMC clouds (see Sect.~\ref{Sec:res}), as well as by comparable CNO elemental abundances measured in different SMC regions \citep[e.g.,][]{1998RMxAC...7..202K,2022MNRAS.517.4497D}. The unusually low $^{18}$O/$^{17}$O isotope ratio detected in LIRS~36 provides a first indication that this value could be characteristic of the SMC, but establishing whether it is representative on a global scale will require future studies with larger samples.

In contrast, the outer Galaxy of the Milky Way, while also metal-poor, is characterized by higher $^{18}$O/$^{17}$O ratios of $\gtrsim$4
\citep[e.g.,][]{2008A&A...487..237W,2020ApJS..249....6Z,2023MNRAS.522..559O}. This difference supports our interpretation that the measured oxygen isotope ratios provide independent evidence for a top-light IGIMF in the SMC. Such a top-light IGIMF may be a common feature for the population of low-metallicity dwarf galaxies. A potential cause for the low star formation efficiencies observed in nearby extremely metal-poor galaxies \citep{2014Natur.514..335S} could be the top-light IGIMF implied by our SMC $^{18}$O/$^{17}$O measurement.

\section{Summary and conclusion} \label{Sec:sum}
In this work, we performed $^{12}$CO $J=2\to 1$, $^{13}$CO $J=2\to 1$, C$^{18}$O $J=2\to 1$, and C$^{17}$O $J=2\to 1$ observations with the Atacama Compact Array (ACA) toward LIRS~36, an active star-forming region in the Small Magellanic Cloud (SMC). Our observations resulted in the first detection of C$^{17}$O in the SMC, enabling a direct and accurate determination of the oxygen isotope ratios: [$^{13}$C/$^{18}$O] = 109.3$\pm$35.6, [$^{13}$C/$^{17}$O] = 95.2$\pm$31.4, [$^{18}$O/$^{17}$O] = 0.87$\pm$0.26, [$^{16}$O/$^{18}$O]$\gtrsim$444, and [$^{16}$O/$^{17}$O]$\gtrsim$386. This establishes a key benchmark for the oxygen isotopic ratio in the SMC and similar metal-poor dwarf galaxies. The $^{18}$O/$^{17}$O ratio in LIRS~36 is lower than all previously reported values for molecular clouds in both extragalactic sources and the Milky Way, indicating chemical enrichment dominated by low-mass stars. This is consistent with the top-light IGIMF expected from the SMC’s long-term low star formation rate. Our findings imply that such a top-light IGIMF may be a general feature of low-metallicity dwarf galaxies. Future systematic surveys of oxygen isotope ratios in local dwarf galaxies and low-metallicity galaxies at high redshift will be essential for advancing our understanding of stellar population and chemical evolution across cosmic time.

%% Please use the acknowledgment and contribution environments. This will 
%% be anonomyized when the "anonymous" style option is used. 
\begin{acknowledgments}
We thank the ALMA staff for their invaluable assistance throughout our observations. Y.G. was supported by the Ministry of Science and Technology of China under the National Key R\&D Program under grant No. 2023YFA1608200, the National Natural Science Foundation of China (NSFC) under grant No. 12427901, and the Strategic Priority Research Program of the Chinese Academy of Sciences under grant No. XDB0800301. We acknowledge the support of the NSFC under grant No. 12041305, 12173016, 1257030642, and 12533003. C.H. acknowledges support by the Chinese Academy of Sciences President's International Fellowship Initiative 
under grant No. 2025PVA0048. C.-H.R. C. acknowledges support from the Deutsches Zentrum für Luft- und Raumfahrt (DLR) grant NS1 under contract no. 50 OR 2214. W.Y. acknowledges the support from the NSFC (12403027), China Postdoctoral Science Foundation (2024M751376), and Jiangsu Funding Program for Excellent Postdoctoral Talent (2024ZB347). X.D.T. acknowledges the support of the National Key R\&D Program of China under grant No. 2023YFA1608002, the Tianshan Talent Training Program of Xinjiang Uygur Autonomous Region under grant No. 2022TSYCLJ0005, and the Chinese Academy of Sciences (CAS) ``Light of West China" Program under grant No. xbzg-zdsys-202212. We thank James Urquhart for his careful reading of the manuscript and for providing valuable comments. We thank Xiaoting Fu for fruitful discussions on stellar nucleosynthesis. This research has made use of NASA's Astrophysics Data System. This paper makes use of the following ALMA data: ADS/JAO.ALMA\#2023.1.01576.S and ADS/JAO.ALMA\#2015.1.00196.S. ALMA is a partnership of ESO (representing its member states), NSF (USA) and NINS (Japan), together with NRC (Canada), MOST and ASIAA (Taiwan), and KASI (Republic of Korea), in cooperation with the Republic of Chile. The Joint ALMA Observatory is operated by ESO, AUI/NRAO and NAOJ. 
\end{acknowledgments}

\begin{contribution}
%%This section gives authors the space to recognize author contributions. The text inside this environment is NOT counted toward the total word quanta. At a minimum, manuscripts are expected to include this text:

%All authors contributed equally to the Terra Mater collaboration.

%% But authors are expected to provide more specific details, e.g. 
%%
Y.G. served as the Principal Investigator of the ACA observing project, performed the data reduction, and drafted the initial manuscript.
Z.Y.Z. first proposed the IGIMF concept in connection with isotope ratios and substantially revised the manuscript.
C.H. contributed to the initial development of the project and to the writing of the manuscript.
C.-H.R.C. assisted in selecting the target from her SMC CO survey.
W.Y. resolved LaTeX issues and prepared the tables.
All authors contributed to the discussion of the results and the revision of the manuscript.
%%OTS obtained the funding and edited the manuscript.
%%EBF provided the formal analysis and validation. He also edited the manuscript.
%%GEH Supervised the undergraduates, wrote the software and administers the project github and Zenodo repositories.
%%
%% Authors can use the Contributor Role Taxonomy (CRediT) at
%% https://credit.niso.org
%% for ideas on how write a good statement tailored to their needs.

\end{contribution}

%% To help institutions obtain information on the effectiveness of their 
%% telescopes the AAS Journals has created a group of keywords for telescope 
%% facilities.
%
%% Following the acknowledgments section, use the following syntax and the
%% \facility{} or \facilities{} macros to list the keywords of facilities used 
%% in the research for the paper.  Each keyword is check against the master 
%% list during copy editing.  Individual instruments can be provided in 
%% parentheses, after the keyword, but they are not verified.
\facilities{ALMA, ACA, TP}

%% Similar to \facility{}, there is the optional \software command to allow 
%% authors a place to specify which programs were used during the creation of 
%% the manuscript. Authors should list each code and include either a
%% citation or url to the code inside ()s when available.
\software{astropy \citep{2022ApJ...935..167A}, 
          NumPy \citep{5725236}, SciPy \citep{jones2001scipy}, 
          Matplotlib \citep{Hunter:2007},
          CASA \citep{2022PASP..134k4501C},
          GILDAS \citep{2005sf2a.conf..721P}, 
          emcee \citep{2013PASP..125..306F}.
          }

%% Appendix material should be preceded with a single \appendix command.
%% There should be a \section command for each appendix. Mark appendix
%% subsections with the same markup you use in the main body of the paper.
%%
%% Each Appendix (indicated with \section) will be lettered A, B, C, etc.
%% The equation counter will reset when it encounters the \appendix
%% command and will number appendix equations (A1), (A2), etc. The
%% Figure and Table counter will not reset.

\appendix
\restartappendixnumbering
%\section{Spectral fitting}
%In this section, we present Gaussian fits to the observed $J=2\to 1$ spectra of $^{12}$CO, $^{13}$CO, and C$^{18}$O, and a hyperfine-structure fit to the C$^{17}$O $J=2\to 1$ spectrum (Fig.~\ref{Fig:fitting}).

%\begin{figure*}[!htbp]
%\centering
%\includegraphics[width = 0.45 \textwidth]{fig/smc-fitting.pdf}
%\caption{{Similar to Fig.~\ref{Fig:N12A}e--\ref{Fig:N12A}h but with Gaussian and hyperfine-structure fitting results overlaid as red dashed lines.}\label{Fig:fitting}}
%\end{figure*}

\section{Short spacing}\label{app.a}
\begin{figure*}[!htbp]
\centering
\includegraphics[width = 0.95 \textwidth]{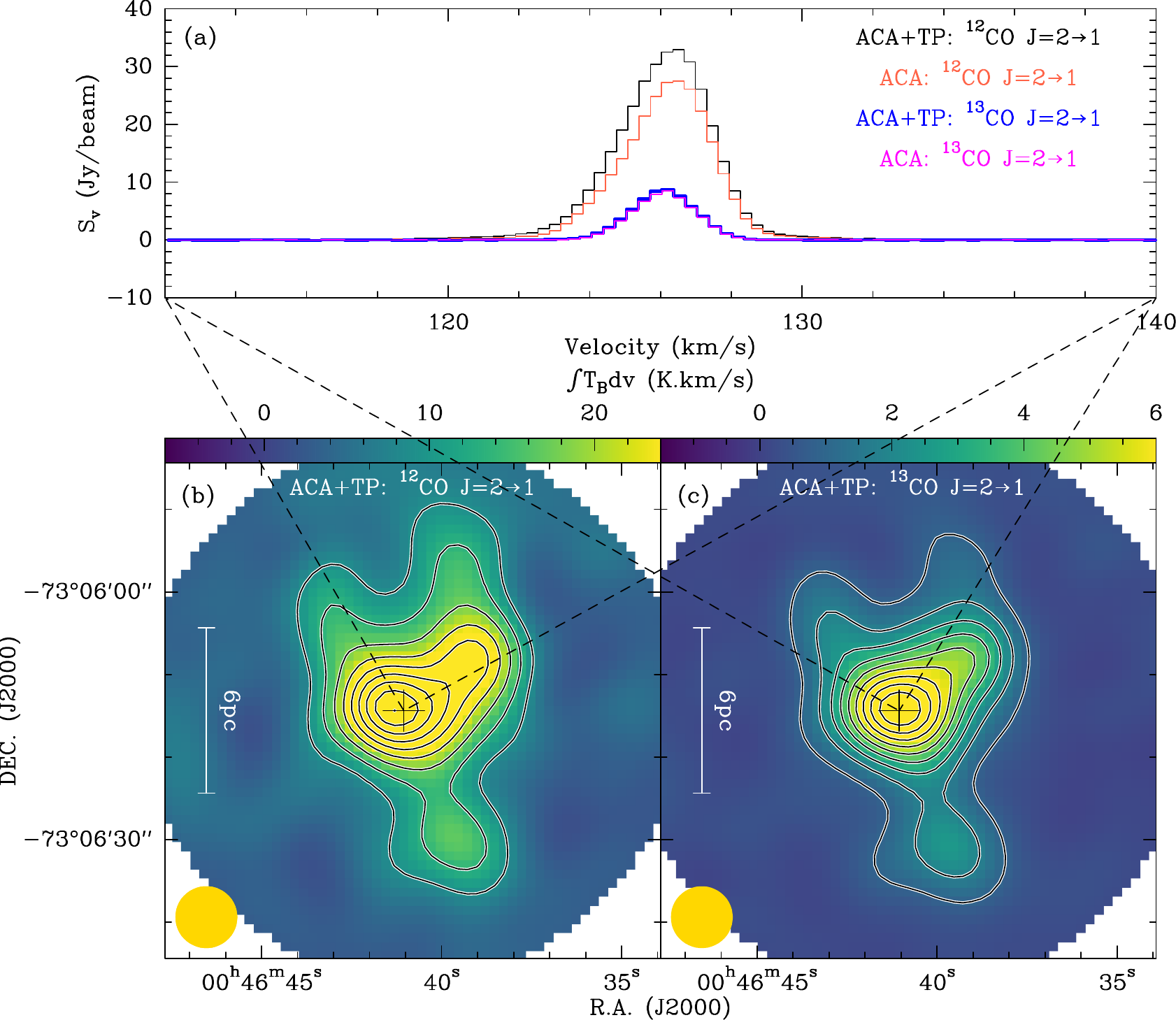}
\caption{{Comparison of the ACA-only and ACA+TP datasets toward LIRS~36. (a) Observed spectra of $^{12}$CO $J=2 \to 1$ and $^{13}$CO $J=2 \to 1$ toward the $^{13}$CO $J=2 \to 1$ peak emission, indicated by the plus signs in panels (b) and (c). (b) Intensity map of $^{12}$CO $J=2\to 1$ integrated from 105~\kms\,to 145~\kms. Contours start at 9.0~K~\kms\,and increase by 4.5~K~\kms. (c) Intensity map of $^{13}$CO $J=2\to 1$ integrated from 122~\kms\,to 129~\kms. Contours start at 0.89~K~\kms\, and increase by 0.89~K~\kms. In panels (b) and (c), the synthesized beam is shown in the lower left corner.}\label{Fig:N12A-mf}}
\end{figure*}

To quantify the contribution of single-dish data in recovering the flux densities of our ACA observations, we compare the ACA-only and ACA+TP datasets in Fig.~\ref{Fig:N12A-mf}. For the total flux density integrated over the field of view, the ACA-only data account for only 70\% and 82\% of the total flux densities in the ACA+TP combined data for $^{12}$CO $J=2 \to 1$ and $^{13}$CO $J=2 \to 1$, respectively. Based on the spectra at the peak position (see Fig.~\ref{Fig:N12A-mf}a), the ACA-only spectra account for 83\% and 97\% of the flux densities of the ACA+TP combined spectra for $^{12}$CO $J=2 \to 1$ and $^{13}$CO $J=2 \to 1$, respectively. This comparison indicates that the $^{12}$CO $J=2 \to 1$ emission exhibits a moderate missing flux, emphasizing the need for total power data to accurately determine its brightness temperatures. In contrast, $^{13}$CO $J=2 \to 1$ emission is more compact and less affected by the absence of short spacings. Notably, the flux density of $^{13}$CO $J=2 \to 1$ shows negligible missing flux at the peak position. Given their lower optical depths, C$^{18}$O $J=2 \to 1$ and C$^{17}$O $J=2 \to 1$ are expected to have spatial distributions more similar to $^{13}$CO $J=2 \to 1$ than to $^{12}$CO $J=2 \to 1$, and are in fact observed to be even more compact (see Figs.~\ref{Fig:N12A}b--\ref{Fig:N12A}d). Therefore, we conclude that the missing flux in the ACA-only data is negligible for the C$^{18}$O $J=2 \to 1$ and C$^{17}$O $J=2 \to 1$ spectra at the peak position, thus supporting the robustness of our analysis in Sect.~\ref{Sec:res}.

\section{Local thermodynamic equilibrium}\label{app.lte}
Under local thermodynamic equilibrium (LTE) conditions, excitation temperatures are indispensable to derive molecular column densities. According to the radiative transfer equation \citep{2015PASP..127..266M}, the peak brightness temperature can be expressed as:
\begin{equation}\label{f.rad}
T_{\rm p} = f[J_{\nu}(T_{\rm ex})-J_{\nu}(T_{\rm bg})][1-{\rm exp}(-\tau)] \;,
\end{equation}
where $J_{\nu}(T)$ is given by:
\begin{equation}\label{f.jv}
J_{\nu}(T) = \frac{h\nu/k}{{\rm exp}(\frac{h\nu}{kT})-1}  \;,
\end{equation}
with $f$ representing the beam dilution factor, $\tau$ the optical depth, $T_{\rm bg}$ the background temperature, which is set to be 2.73~K \citep{2009ApJ...707..916F}. Here, the Planck constant, $h$, is 6.626$\times 10^{-27}$ erg~s, the Boltzmann constant, $k$, is 1.38$\times 10^{-16}$ erg~K$^{-1}$, and $\nu$ is the rest frequency. Assuming that $^{12}$CO $J=2\to 1$ is optically thick and neglecting the beam dilution effects (i.e., $\tau \approx \infty$ and $f\sim 1$),  we can derive excitation temperatures from the peak brightness temperatures of $^{12}$CO $J=2\to 1$ using Eqs.~(\ref{f.rad})--(\ref{f.jv}). Assuming that the $J=2\to 1$ transitions of $^{13}$CO, C$^{18}$O, and C$^{17}$O have the same excitation temperature as that of $^{12}$CO $J=2 \to 1$, their optical depths can be estimated with Eqs.~(\ref{f.rad}).

To model the excitation temperature and molecular column densities simultaneously, we use the CLASS extension Weeds for the LTE spectral modeling \citep{2011A&A...526A..47M}. For the synthetic Weeds model, we assume a single Gaussian component for all molecules for simplicity. The Weeds model is defined by a set of five parameters: molecular column density, excitation temperature, source size, systemic velocity, and FWHM line width. The systemic velocities and FWHM line widths are based on Table~\ref{Tab:obs}. The FWHM line width of C$^{17}$O $J=2\to 1$ is set to be identical to C$^{18}$O $J=2\to 1$, as the FWHM line width of C$^{17}$O $J=2\to 1$ is overestimated by the Gaussian fitting in Table~\ref{Tab:obs} as a result of the hyperfine-structure splitting. Therefore, we have three free parameters for each line. In this study, we only aim to derive the column densities for $^{13}$CO, C$^{18}$O, and C$^{17}$O. 

To fit the observed spectra, we employed the \textit{emcee} package \citep{2013PASP..125..306F}, which applies the affine-invariant ensemble sampler \citep{2010CAMCS...5...65G} within a Markov chain Monte Carlo (MCMC) framework. This method enables a thorough exploration of the posterior distributions of the model parameters. We assumed uniform priors for the excitation temperature, $T_{\rm ex}$, $^{13}$CO column density, $N_{\rm ^{13}CO}$, C$^{18}$O column density, $N_{\rm C^{18}O}$, and C$^{17}$O column density, $N_{\rm C^{17}O}$. The posterior probability was calculated as the product of the prior and the likelihood, where the likelihood was taken to be Gaussian, $\mathcal{L} \propto \exp(-\chi^{2}/2)$, with 
\begin{equation}\label{f.chi}
\chi^{2} = \sum_{i} \left( \frac{T_{{\rm obs},i} - T_{{\rm mod},i}}{\sigma_{i}} \right)^{2} \;,
\end{equation}
and $T_{{\rm obs},i}$, $T_{{\rm mod},i}$, and $\sigma_{i}$ denoting the observed and modeled peak brightness temperature, as well as the corresponding 1$\sigma$ observational uncertainties, respectively. The MCMC sampling was carried out with 10 walkers over 4500 iterations, following an initial burn-in phase to ensure convergence. The reported best-fit values and associated 1$\sigma$ errors correspond to the 16th and 84th percentiles of the posterior distributions.

\begin{figure*}[!htbp]
\centering
\includegraphics[width = 0.95 \textwidth]{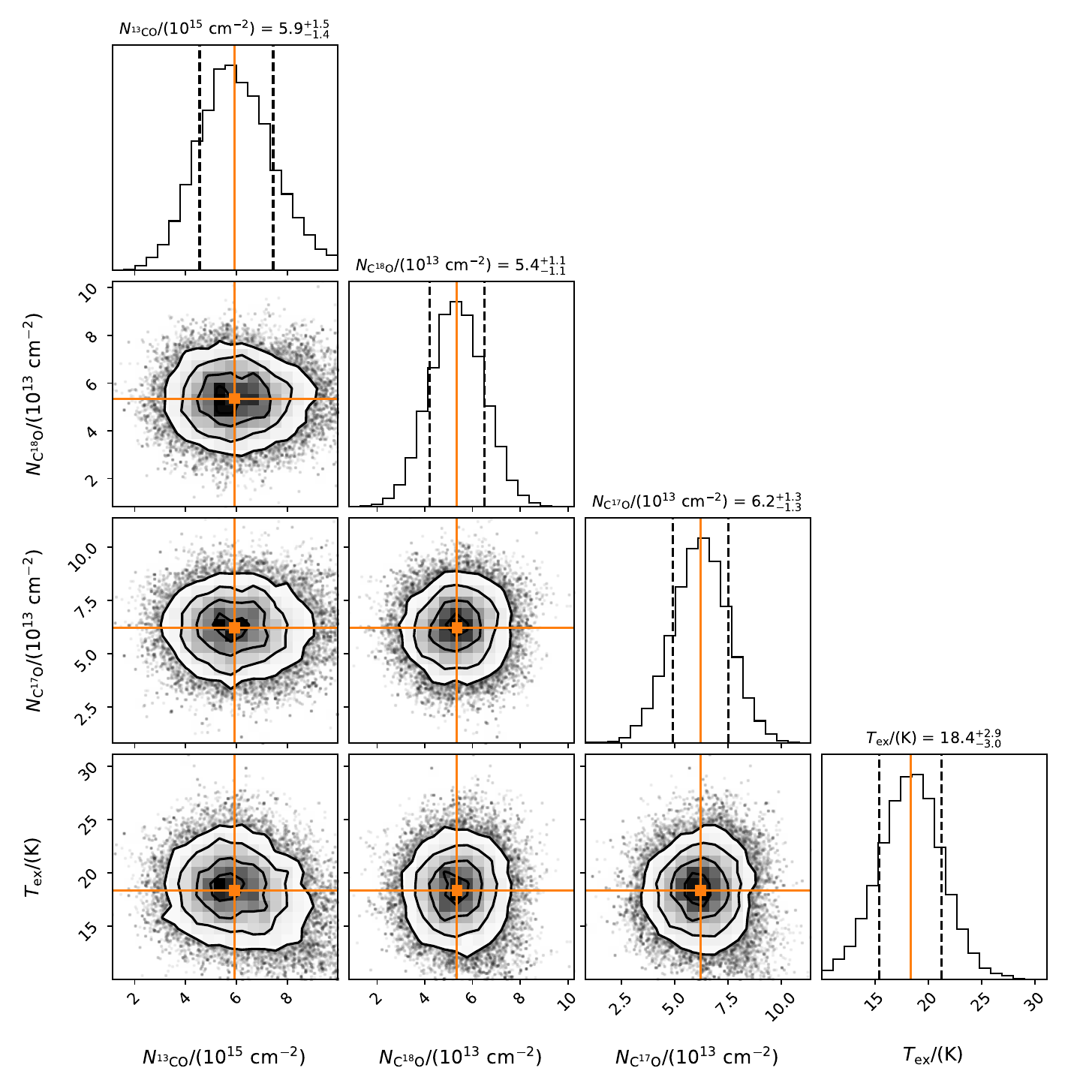}
\caption{{Posterior probability distributions of $^{13}$CO column densities, C$^{18}$O column densities, C$^{17}$O column densities, and excitation temperatures for the Weeds LTE modeling, with the maximum posterior possibility point in the parameter space highlighted by orange lines and points. Contours represent the 0.5, 1.0, 1.5, and 2.0$\sigma$ confidence intervals. The vertical dashed black lines represent the 1$\sigma$ spread. }\label{Fig:mcmc}}
\end{figure*}

Assuming the four lines trace the same parcel of molecular gas, we assume the same source size for the four lines. The source size is directly related to the beam dilution factor with $f=\frac{\theta_{\rm s}^{2}}{\theta_{\rm s}^{2}+\theta_{\rm b}^{2}}$. For the fiducial case, we neglect the beam dilution effects by assuming $f\sim 1$. The modeling results are shown in Fig.~\ref{Fig:mcmc}, where the four parameters are well constrained. %The best-fit synthetic Weeds spectra are shown in red in Figs.~\ref{Fig:N12A}f--\ref{Fig:N12A}h, which match the observed spectra excellently. 
Our approach results in $T_{\rm ex}$=18.4$\pm$3.0~K, $N_{^{13}{\rm CO}}$=(5.9$\pm$1.5)$\times 10^{15}$~cm$^{-2}$, $N_{{\rm C^{18}O}}$=(5.4$\pm$1.1)$\times 10^{13}$~cm$^{-2}$, and $N_{{\rm C^{17}O}}$=(6.2$\pm$1.3)$\times 10^{13}$~cm$^{-2}$. The derived isotope ratios are [$^{13}$C/$^{18}$O] = 109.3$\pm$35.6, [$^{13}$C/$^{17}$O] = 95.2$\pm$31.4, and [$^{18}$O/$^{17}$O] = 0.87$\pm$0.26. In order to test the impact of the assumed source size on the derived isotope ratios, the MCMC approach was performed again by assuming $\theta_{\rm s}$=5\rlap{.}\arcsec4\,which was inferred from the Gaussian fitting of the compact component in Fig.~\ref{Fig:N12A}d. The physical parameters change to $T_{\rm ex}$=46.5$\pm$8.2~K, $N_{^{13}{\rm CO}}$=(2.5$\pm$0.6)$\times 10^{16}$~cm$^{-2}$, $N_{{\rm C^{18}O}}$=(2.2$\pm$0.5)$\times 10^{14}$~cm$^{-2}$, and $N_{{\rm C^{17}O}}$=(2.5$\pm$0.6)$\times 10^{14}$~cm$^{-2}$, indicating [$^{13}$C/$^{18}$O] = 113.3$\pm$37.6, [$^{13}$C/$^{17}$O] = 100.0$\pm$33.9, and [$^{18}$O/$^{17}$O] = 0.88$\pm$0.29. This suggests that the different source sizes affect $T_{\rm ex}$, $N_{\rm ^{13}CO}$, $N_{\rm C^{18}O}$, and $N_{\rm C^{17}O}$ significantly, but the isotope ratios remain almost identical. Therefore, the derived isotope ratios in the SMC are robust.  

\bibliography{smc}{}
\bibliographystyle{aasjournalv7}

%% This command is needed to show the entire author+affiliation list when
%% the collaboration and author truncation commands are used.  It has to
%% go at the end of the manuscript.
%\allauthors

%% Include this line if you are using the \added, \replaced, \deleted
%% commands to see a summary list of all changes at the end of the article.
%\listofchanges

\end{document}